\documentclass{ws-rv9x6}

\def\nablab{\mbox{\boldmath $\nabla$}}

\begin{document}

\setcounter{chapter}{0}

\chapter{Field Theoretical Approaches to the Superconducting
Phase Transition}

\markboth{F.S. Nogueira, H. Kleinert}{Field Theoretical Approaches to the Superconducting
Phase Transition}

\author{Flavio S. Nogueira and Hagen Kleinert}

\address{Institut f\"ur Theoretische Physik,
Freie Universit\"at Berlin, Arnimallee 14, D-14195 Berlin, Germany\\
E-mail: nogueira@physik.fu-berlin.de}

\begin{abstract}
Several field theoretical approaches to the superconducting phase
transition are discussed. Emphasis is given to theories of scaling
and renormalization group in the context of the Ginzburg-Landau
theory and its variants. Also discussed is the duality approach,
which allows to access the strong-coupling limit of the
Ginzburg-Landau theory.
\end{abstract}

\section{Introduction}

The Ginzburg-Landau (GL) \index{Ginzburg-Landau theory} 
phenomenological theory of
superconductivity\cite{GL} was proposed long before the famous
Bardeen-Cooper-Schrieffer (BCS) \index{BCS theory}
microscopic theory of
superconductivity.\cite{BCS} A few years after the BCS theory,
Gorkov\cite{Gorkov}
derived the GL theory from the BCS theory. This has been
done by deriving an effective theory for the Cooper pairs valid
in the neighborhood of the critical temperature $T_c$.
The modern derivation
proceeds most elegantly
via functional integrals, introducing
a collective quantum field  $ \Delta ({\bf x},t)$
for the Cooper pairs via a Hubbard-Stratonovich
transformation and integrating out the fermions
in the partition function.\cite{CQF}

Amazingly, the GL theory has kept
great actuality up to now. It is highly relevant
for the description of high-$T_c$ superconductors,
even though
the original BCS theory is inadequate to treat
these materials.
The success of the GL theory in the study of
modern problems of superconductivity lies on its
universal effective character,
 the details of the microscopic model being unimportant.
In the neighborhood of
 the critical point
the GL theory possesses a wide range of application, many of them
 outside the field of the
superconductivity. Perhaps, the most famous example in
condensed matter physics is application to
 the smectic-A to nematic phase transition in
liquid crystals.\cite{deGennes} 
In elementary particle physics,
Gorkov's derivation
of the GL theory has been imitated
 starting out from a four-dimensional relativistic version of
the BCS model, the so-called Nambu-Jona-Lasinio model.
The GL field in the resulting GL theory
describes quark-antiquark bound states, the mesons
 $ \pi $, $ \sigma $, $\rho$, and  $A_1$.\cite{HQT} 
This model  is still of wide use
in nuclear physics.
Another relativistic GL model is needed to
 generate the masses of the vector bosons
 $W$ and $Z$
in the unified theory of electromagnetic and weak interactions
in a renormalizable way.\cite{ZJ}
This is done by a nonabelian analog of the {\em Meissner effect\/},
which
 in this context
 is called
{\em Higgs mechanism\/}.

In 1973 Coleman and Weinberg\cite{CW} showed that
the four-dimensional abelian
 GL model (or scalar electrodynamics,
in the language of elementary particle physics) 
exhibits a spontaneous
mass generation in an initially massless theory.
In the language of statistical mechanics,
this implies a first-order transition
if the  mass of the scalar field
passes through zero.
One year later, Halperin,
Lubensky, and Ma (HLM) observed a similar phenomenon
in the three-dimensional GL theory
 of superconductivity.\cite{HLM} These
papers inaugurated a new era in the study of the GL model. It was the
first time that renormalization group (RG) methods were used to
study the superconducting phase transition.

At the mean-field level, the GL model exhibits a second-order
phase transition. The HLM analysis, however, concludes that
fluctuations change the order of the transition to first.
The phase transition would be of second-order only if the
number of complex components of the order
parameter is absurdly high. If the
number of complex components is $N/2$, the one-loop RG analysis
of HLM gave the
lower bound
$N>N_c=365$
 for a second-order transition.
No infrared stable charged fixed point was found for
$N\leq N_c$. However, years later Dasgupta and Halperin\cite{Dasgupta}
raised doubts on this result
by using duality arguments and a Monte Carlo simulation
of a lattice model in the London
limit, to demonstrate
that the transition for $N=2$ in the type II regime
is of second order. The RG result that the phase transition
is always of first order seems therefore to be an artifact of the
$\epsilon$-expansion. Shortly after this,
 Kleinert\cite{Kleinert} performed
a quantitative duality on the lattice
and found a disorder field theory\cite{DO}
which clarified the
discrepancy between the HLM RG result and the numerical simulations
of Dasgupta and Halperin. He showed that there exists a tricritical
point in the phase diagram of the superconductor
on the separation line
between first- and second-order regimes. This result and
the numerical prediction\cite{Kleinert} were fully confirmed only recently
by large scale Monte Carlo simulations of Mo {\it et al.}\cite{SUD}

In the last ten years the RG approach to the GL model was revisited
by several groups.\cite{Kiometzis,Folk,Berg,Herbut,KleinNog1}
The aim was to improve the RG analysis
in such a way as to obtain the charged fixed point for $N=2$
 predicted by the duality scenario.
Despite considerable progress,
 our understanding of this problem is far from satisfactory.

In this paper we shall review the
basic  field theoretic
ideas relevant for  the understanding
of the superconducting phase transition. The
reader is supposed to have some familiarity
with RG theory and duality
transformations. For a recent review on the RG approach to the GL 
model which focus more on resummation of $\epsilon$-expansion 
results, including calculations of amplitude ratios, see 
Ref.~\refcite{Yurij}. 

\section{Review of the HLM theory \index{Halperin-Lubensky-Ma theory}}

The GL Hamiltonian is given by

\begin{equation}
\label{GL}
{\cal H}=\frac{1}{2}({\mbox{\boldmath $\nabla$}}\times{\bf
A})^2+|({\mbox{\boldmath $\nabla$}}-ie{\bf
A})\psi|^2+m^2|\psi|^2+\frac{u}{2}|\psi|^4, 
\end{equation}
where $\psi$ is the order field and ${\bf A}$ is the fluctuating 
vector potential. 
The partition function is written in the form of a functional integral as

\begin{equation}
\label{partition}
Z=\int{\cal D}{\bf A}{\cal D}\psi^{\dag}{\cal D}\psi
\det(-\nablab^2)\delta(\nablab\cdot{\bf A})\exp\left(-\int d^3 r {\cal H}\right),
\end{equation}
where
a  delta
functional
in the measure of integration enforces  the Coulomb gauge
$\nablab\cdot{\bf A}=0$.
The factor $\det(-\nablab^2)$ is the associated Faddeev-Popov
determinant.\cite{ZJ}
The above Hamiltonian coincides with
the euclidian version of
the Abelian Higgs model in particle physics. The Coulomb gauge is
the euclidian counterpart of the relativistic Lorentz gauge. 
The Faddeev-Popov determinant should be included in order 
cancel a contribution $1/\det(-\nablab^2)$ that arises 
upon integration over ${\bf A}$ taking into account the 
constraint $\nablab\cdot{\bf A}=0$.\cite{ZJ,KleinertBook} 

\subsection{HLM mean-field theory}

Let us make the following change of variables in the partition
function (\ref{partition}):

\begin{equation}
\psi=\frac{1}{\sqrt{2}}\rho ~e^{i\theta},
~~~~~{\bf A}={\bf a}+\frac{1}{e}\nablab\theta,
\end{equation}
which brings the  partition function
to the form
\begin{equation}
\label{partition1}
Z=\int{\cal D}{\bf a}~{\cal D}\rho~{\cal D}\theta
~\rho~\det(-\nablab^2)\delta(\nablab\cdot{\bf a}+e^{-1}\nablab^2\theta)\exp
\left(-\int d^3 r {\cal H}\right),
\end{equation}
with
\begin{equation}
{\cal H}=\frac{1}{2}({\mbox{\boldmath $\nabla$}}\times{\bf
a})^2+\frac{e^2}{2}\rho^2{\bf a}^2+\frac{1}{2}(\nablab\rho)^2
+\frac{m^2}{2}\rho^2+\frac{u}{8}\rho^4.  \label{GLrho}
\end{equation}
In these variables, the
Hamiltonian does not depend on $\theta$.
This
 change of variables is allowed only in a region
where the system has few vortex lines,
since otherwise
the cyclic  nature of the  $\theta$-field becomes
relevant,
and
$\nablab\, e^{i\theta}$ is no longer equal to
$i\nablab\theta\, e^{i\theta}$, as assumed in going
from (\ref{GL})
to (\ref{GLrho}), but equal to
$i(\nablab\theta-2\pi {\bf n}) e^{i\theta}$, where ${\bf n}$ is a
vortex gauge field. \cite{DO,cam}
  This problem of the Hamiltonian (\ref{GLrho})
which is said to be in the {\em unitary gauge\/}
will need special attention.

Now we can use the delta function to integrate out $\theta$.
The result of this integration cancels out the Faddeev-Popov
determinant.

Next we assume that $\rho$ is uniform, say $\rho=\bar{\rho}={\rm const}$.
Since the Hamiltonian is quadratic in ${\bf a}$, the
${\bf a}$ integration can be done straightforwardly to obtain
the free energy density:

\begin{equation}
\label{FE}
{\cal F}=\frac{1}{2V}{\rm Tr}\ln[(-\nablab^2+e^2\bar{\rho}^2)
\delta_{\mu\nu}+\partial_\mu\partial_\nu]
-\frac{1}{2V}\delta^3(0){\rm Tr}\ln(e^2\bar{\rho}^2)+\frac{m^2}{2}\bar{\rho}^2
+\frac{u}{8}\bar{\rho}^4,
\end{equation}
where $V$ is the (infinite) volume and the term
$\delta^3(0){\rm Tr}\ln(e^2\bar{\rho}^2)/2V$
comes from the exponentiation of the functional Jacobian in
Eq. (\ref{partition1}).
In order to evaluate
the ${\rm Tr}\ln$ in Eq. (\ref{FE})
we have to use the Fourier transform of
the operator

\begin{equation}
M({\bf r},{\bf r}')=[(-\nablab^2+e^2\bar{\rho}^2)
\delta_{\mu\nu}+\partial_\mu\partial_\nu]\delta
^3({\bf r}-{\bf r}').
\end{equation}
The operator $M({\bf r},{\bf r}')$ is diagonal in momentum
space. Indeed, we can write

\begin{equation}
M({\bf r},{\bf r}')=\int\frac{d^{3} p}{(2\pi)^3}
e^{i{\bf p}\cdot({\bf r}-{\bf r}')}\hat{M}({\bf p}),
\end{equation}

\begin{equation}
\label{Mftr}
\hat{M}({\bf p})=({\bf p}^2+e^2\bar{\rho}^2)P^T_{\mu\nu}({\bf p})
+e^2\bar{\rho}^2 P^L_{\mu\nu}({\bf p}),
\end{equation}
where $P^T_{\mu\nu}({\bf p})=\delta_{\mu\nu}-p_\mu p_\nu/{\bf p}^2$
and $P^L_{\mu\nu}({\bf p})=p_\mu p_\nu/{\bf p}^2$
are the transverse and
longitudinal projectors, respectively.
 Now, the
${\rm Tr}\ln M({\bf r},{\bf r}')$ is obtained first by taking
the logarithm of the transversal and longitudinal parts of
$\hat{M}({\bf p})$ and tracing over the vector indices. The second
step is to trace out the coordinates by integrating over
${\bf r}$ for ${\bf r}={\bf r}'$. This produces an overall volume
factor $V$ which cancels out exactly the $1/V$ factor
in Eq. (\ref{FE}). Note that the term proportional to
$\delta^3(0)\ln(e^2\bar{\rho}^2)$ drops out in the calculations.
If we write $\bar{\rho}=|\bar{\psi}|$,
the end result is the celebrated HLM mean field free energy:

\begin{equation}
\label{HLMfe}
{\cal F}_{\rm HLM}=-\frac{e^3|\bar{\psi}|^3}{6\pi}
+\frac{m^2}{2}|\bar{\psi}|^2+\frac{u}{8}|\bar{\psi}|^4.
\end{equation}
Due to the cubic term in Eq. (\ref{HLMfe}), the transition is
of first order.
The basic problem 
with this argument was pointed out in Ref.~\refcite{Kleinert}. 
In the critical regime of a  type-II superconductor.
the order field
contains numerous lines of zeros which
make it impossible to use uniformity
assumption
$\rho=\bar{\rho}={\rm const}$.

\subsection{Renormalization group in $d=4-\epsilon$ dimensions 
\index{Renormalization group}}

In the original HLM work, the RG calculations were performed
using the Wilson version of the renormalization group,\cite{Wilson} where
fast modes are integrated out to obtain an effective theory in terms
of the slow modes. The field theoretical approach using the
Callan-Symanzik equation gives an equivalent result in a perturbative
setting. We shall discuss the RG calculation in $d=4-\epsilon$
dimensions in this context.

The dimensionless renormalized couplings are defined by

\begin{equation}
\label{couplings}
f\equiv\mu^{-\epsilon}Z_A e_0^2,
~~~~~~~~~g\equiv\mu^{-\epsilon}Z_\psi^2 Z_g^{-1} u_0,
\end{equation}
where $\mu$ gives the mass scale of the problem. 
Note that in Eq. (\ref{couplings}) we have denoted the bare couplings
by a zero subindex. The bare fields are denoted by $\psi_0$ and 
${\bf A}_0$. Accordingly, the bare mass is denoted by $m_0$.  
We shall use this notation from now on. 
 
The renormalization constants are defined such that
the ``renormalized Hamiltonian'' ${\cal H}_r$ is given by the following
rewriting of the bare Hamiltonian:

\begin{equation}
\label{Hr}
{\cal H}_r({\bf A},\psi;m^2,u,e)=
{\cal H}(Z_A^{1/2}{\bf A},Z_\psi^{1/2}\psi;Z_m Z_\psi^{-1}
m^2,Z_g Z_\psi^{-2}u,Z_A^{-1/2}e).
\end{equation}
From Eq. (\ref{Hr}) we see that the renormalized 
fields are given by $\psi=Z_\psi^{-1/2}\psi_0$ and 
${\bf A}=Z_A^{-1/2}{\bf A}_0$. 

The calculations are more easily done if we set $m=0$ and evaluate
the Feynman graphs at nonzero external momenta to avoid infrared
divergences. The external momenta in a four leg graph will be
taken at the symmetry point:

\begin{equation}
\label{SP}
{\bf p}_i\cdot{\bf p}_j=\frac{\mu^2}{4}(4\delta_{ij}-1).
\end{equation}

\begin{figure}
\centering
\includegraphics[width=2cm,angle=-90]{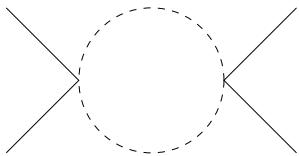}
\caption{Feynman graph contributing to the
$f^2$ term in $\beta_g$. The dashed lines
represent the vector potential propagator, while the external lines
represent the order parameter}
\label{Fish}
\end{figure}

Even if $m\neq 0$ we have to face severe infrared divergences in this
problem. This is due to the masslessness of the vector potential
field. The free ${\bf A}$ propagator is given in the Coulomb
gauge by

\begin{equation}
\label{Aprop}
D_{\mu\nu}({\bf p})=\frac{1}{{\bf p}^2}\left(\delta_{\mu\nu}
-\frac{p_\mu p_\nu}{{\bf p}^2}\right).
\end{equation}
Thus, the graph in Fig.~\ref{Fish} is proportional to
the integral

\begin{equation}
\left.\int\frac{d^d k}{(2\pi)^d}\frac{1}{{\bf k}^2({\bf p}-{\bf k})^2}
\right|_{\rm SP}=
\frac{\pi^{d/2}}{(2\pi)^d}\frac{\Gamma(2-d/2)\Gamma^2(d/2-1)}{\Gamma(
d-2)}\mu^{d-4},
\end{equation}
evaluated at the symmetry point (SP)
as prescribed in Eq. (\ref{SP}).

The $\beta$-functions \index{$\beta$-functions} are defined by

\begin{equation}
\beta_f\equiv\mu\frac{\partial f}{\partial\mu},
~~~~~~~~~\beta_g\equiv\mu\frac{\partial g}{\partial\mu}.
\end{equation}
These $\beta$-functions are defined for the dimension 
$d$ in the interval $(2,4]$. For $d>4$ the theory is 
no longer renormalizable and the $\beta$-functions are 
not defined. At $d=2$ the infrared divergences are very 
strong and require a special treatment. 
 
\begin{figure}
\centering
\includegraphics[width=6cm,angle=0]{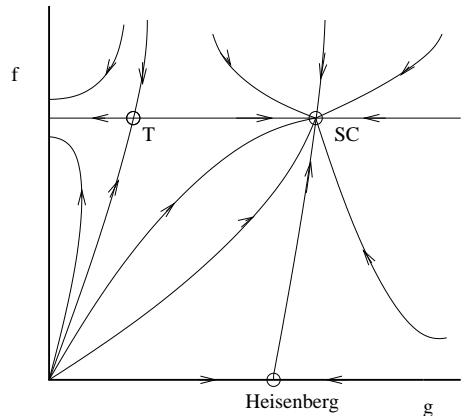}
\caption{Schematic flow diagram for the GL model for 
$N>N_c$.}
\label{Flow}
\end{figure}

Let us assume that the order parameter field has $N/2$ complex
components. Then,
for any dimension $d\in(2,4]$ the $\beta$-functions are given at
one-loop order by

\begin{equation}
\label{betaf1}
\beta_f=(4-d)\{-f+N A(d)f^2\},
\end{equation}

\begin{equation}
\label{betag1}
 \beta_g=(4-d)\bigg\{-g+B(d)
\left[-2(d-1)fg+\frac{N\!+\!8}{2}g^2+
2(d\!-\!1)f^2\right]\bigg\},
\end{equation}
where

\begin{equation}
A(d)=-\frac{\Gamma(1-d/2)\Gamma^2(d/2)}
{(4\pi)^{d/2}\Gamma(d)},
\end{equation}

\begin{equation}
B(d)=\frac{\Gamma(2-d/2)\Gamma^2(d/2-1)}
{(4\pi)^{d/2}\Gamma(d-2)}.
\end{equation}
If we set $d=4-\epsilon$ and expand to first order in
$\epsilon$, we obtain\cite{HLM,Lawrie}

\begin{equation}
\label{betaf2}
\beta_f=-\epsilon f+\frac{N}{48\pi^2}f^2,
\end{equation}

\begin{equation}
\label{betag2}
\beta_g=-\epsilon g-\frac{3fg}{4\pi^2}
+\frac{N+8}{8\pi^2}g^2+\frac{3}{8\pi^2}f^2.
\end{equation}
From Eqs. (\ref{betaf2}) and (\ref{betag2}) we see easily that
{\it charged} fixed points 
\index{Charged fixed points}
exist only if $N>N_c=365.9$. Thus, if
$N$ is large enough to allow for the existence of charged
fixed points we obtain the flow diagram 
\index{Flow diagram} shown schematically
in Fig. \ref{Flow}. In the figure the arrows correspond 
to $\mu\to 0$. 
There are four fixed points. The Gaussian
fixed point is trivial and corresponds to $f_*=g_*=0$. It governs
the ordinary mean-field behavior. There is one non-trivial
uncharged fixed point, labeled ``Heisenberg'' in the figure, which
governs the $N$-component
Heisenberg model  universality class. For $N=2$ the superfluid $^4$He 
belongs to this universality class (in this case we speak of a
$XY$ universality class).
The Heisenberg fixed point is
unstable for non-zero charge.
There are two charged
fixed points. The one labeled $SC$ in the figure is infrared
stable and governs the superconducting phase transition.
Its infrared stability ensures that the phase transition is
second-order. The second charged fixed point is labeled with
a $T$ and is called the tricritical fixed point. It is infrared
stable along the line starting in the Gaussian fixed point and
unstable along the $g$-direction. The line of stability of
the tricritical fixed point \index{Tricritical fixed point} 
is called the tricritical line.
The tricritical line separates the regions in the flow diagram
corresponding to first- and second-order phase transition.

As we shall see later on in these lectures, the flow diagram
shown in Fig. \ref{Flow} should also be valid for $N=2$. 
\footnote{In the context of the $\epsilon$-expansion, see 
for example the two-loops resummed RG analysis by 
Folk and Holovatch. \cite{Folk,Yurij}}
For
the moment let us remark that if we use the $\beta$-functions
in fixed dimension as given in Eqs. (\ref{betaf1}) and (\ref{betag1})
the value of $N_c$ is considerably smaller at $d=3$. Indeed,
by setting $d=3$ in Eqs. (\ref{betaf1}) and (\ref{betag1}), we
obtain that charged fixed points exist for $N>N_c=103.4$.

\subsection{Critical exponents \index{Critical exponents}}

The critical exponents can be evaluated using standard methods.\cite{HKSF}
The $\eta$ exponent  
is obtained as the value of the
RG function

\begin{equation}
\label{gammapsi}
\gamma_\psi=\mu\frac{\partial\ln Z_\psi}{\partial\mu},
\end{equation}
at the infrared stable fixed point. The $\eta$ exponent governs the
large distance behavior of the order field correlation function at the
critical point:

\begin{equation}
\langle\psi({\bf r})\psi^\dag({\bf r}')\rangle\sim
\frac{1}{|{\bf r}-{\bf r}'|^{d-2+\eta}}.
\end{equation}
The critical exponent $\nu$, governing the scaling of the
correlation length $\xi=m^{-1}\sim t^{-\nu}$, with $t$ being the
reduced temperature, is obtained as the infrared 
stable fixed point value of the
RG function $\nu_\psi$ defined by the equation

\begin{equation}
\frac{1}{\nu_\psi}-2=\gamma_m,
\end{equation}
where

\begin{equation}
\gamma_m=\mu\frac{\partial}{\partial\mu}\ln\left(\frac{Z_m}{Z_\psi}
\right).
\end{equation}
In an approach where the correlation functions are computed at
the critical point the mass renormalization $Z_m$ must be computed
through $|\psi|^2$ insertions in the 2-point function.\cite{HKSF,ZJ} 
For any dimension $d\in(2,4]$, we obtain the
one-loop result:

\begin{equation}
\gamma_\psi=(1-d)(4-d)B(d)f,
\end{equation}

\begin{equation}
\label{gammam}
\gamma_m=(N+2)(d-4)B(d)g/2-\gamma_\psi.
\end{equation}
Once the critical exponents $\eta$ and $\nu$ are evaluated, all the
other critical exponents can be obtained using the standard
scaling relations. We shall prove later that the standard scaling
relations apply also in the case of the GL model.

\subsection{$1/N$ expansion \index{$1/N$ expansion}}

The $1/N$ expansion is one of the most popular non-perturbative
methods in the field theoretical and statistical physics literature.
The critical exponents of the GL model at ${\cal O}(1/N)$ were
calculated at $d=3$ by HLM. Since the original paper does not
contain any detail of the calculation, we shall outline it here.
The large $N$ limit in the GL model is taken at $Nu$ and $Ne^2$
fixed.

\begin{figure}
\centering
\includegraphics[width=8cm,angle=0]{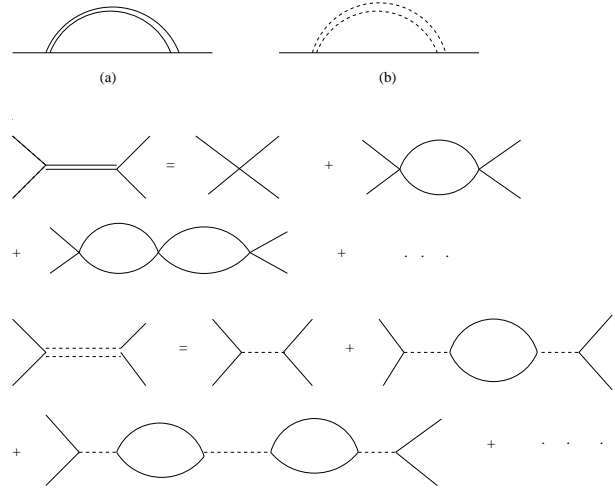}
\caption{Feynman graph contributing to the
momentum dependent part of the self-energy of the
order field propagator. The double lines represent dressed
propagators. The graphs (a) and (b) contain effective
vertices given by an infinite sum of a chain of bubbles.}
\label{KleinF3}
\end{figure}

Let us calculate the
critical exponent $\eta$. The relevant graphs are shown in
Fig. \ref{KleinF3}. The idea is to pick up the
${\bf p}^2\ln|{\bf p}|$ contribution of
$\Gamma^{(2)}({\bf p})=G^{-1}({\bf p})$, where $G$ is the
order field propagator. To keep the theory massless, we
have to subtract the ${\bf p}=0$ contribution of
the self-energy. The subtracted
contribution coming from the graph (a) in Fig. \ref{KleinF3} is

\begin{equation}
\label{se(a)}
\Sigma_a({\bf p})=\int\frac{d^3 k}{(2\pi)^3}V({\bf k})
\left[\frac{1}{({\bf p}-{\bf k})^2}-\frac{1}{{\bf k}^2}\right],
\end{equation}
where

\begin{equation}
\label{V}
V({\bf k})=\frac{2u}{1+uN\Pi({\bf k})},
\end{equation}
and $\Pi({\bf k})$ is the polarization
bubble:

\begin{equation}
\label{bubble}
\Pi({\bf p})=\int\frac{d^3 k}{(2\pi)^3}G_0({\bf p}-{\bf k})G_0({\bf k}),
\end{equation}
where $G_0$ is the free scalar propagator. In the massless case
we have simply

\begin{equation}
\Pi({\bf p})=\frac{1}{8|{\bf p}|}.
\end{equation}
When $e=0$ Eq. (\ref{V}) gives the effective interaction of the
$O(N)$ model\cite{ZJ}. The $-\eta_a{\bf p}^2\ln|{\bf p}|$
contribution from Eq. (\ref{se(a)}) gives

\begin{equation}
\label{etaa}
\eta_a=\frac{8}{3\pi^2N},
\end{equation}
which is just the $\eta$-exponent of the $O(N)$ model
at ${\cal O}(1/N)$.\cite{ZJ}

In order to compute the contribution from the graph (b) of
Fig. (\ref{KleinF3}), we need to calculate the {\it dressed}
vector potential propagator.
The dressed propagator is obtained by summing the chain of
bubbles shown in Fig. \ref{KleinF3}. The result is

\begin{equation}
\label{dress}
D_{\mu\nu}({\bf p})=\frac{1}{{\bf p}^2+\Sigma_A({\bf p})
}\left(\delta_{\mu\nu}-
\frac{p_\mu p_\nu}{{\bf p}^2}\right),
\end{equation}
where in the massless case

\begin{equation}
\Sigma_A({\bf p})=\frac{Ne^2}{32}|{\bf p}|.
\end{equation}
The self-energy contribution from graph (b) is then

\begin{equation}
\Sigma_b({\bf p})=-e^2\int\frac{d^3 k}{(2\pi)^3}\frac{(2p_\mu-k_\mu)
(2p_\nu-k_\nu)}{({\bf p}-{\bf k})^2}D_{\mu\nu}({\bf k}).
\end{equation}
Thus, we obtain the contribution $-\eta_b{\bf p}^2\ln|{\bf p}|$,
where

\begin{equation}
\label{etab}
\eta_b=-\frac{128}{3\pi^2N}
\end{equation}
The $\eta$-exponent of the GL model at ${\cal O}(1/N)$ is obtained
by adding the contributions from Eqs. (\ref{etaa}) and
(\ref{etab}):

\begin{equation}
\label{eta1/N}
\eta=\eta_a+\eta_b=-\frac{40}{\pi^2N}.
\end{equation}

The critical exponent $\nu$ at ${\cal O}(1/N)$ can be 
computed through similar calculations, except that 
one needs to consider massive
propagators for the order field. Instead picking up the
contribution ${\bf p}^2\ln|{\bf p}|$, we take the
$m^2\ln m$ one from the self-energy at zero momentum.
This gives in fact the critical exponent $\gamma$. The
critical exponent $\nu$ is obtained straightforwardly
by using the scaling relation $\gamma=\nu(2-\eta)$.
The end result is:

\begin{equation}
\nu=1-\frac{96}{\pi^2N}.
\end{equation}

Note that in the $1/N$ expansion no tricritical fixed
point is generated at ${\cal O}(1/N)$. The reason for that
comes from the fact that the graph in Fig. \ref{Fish} is
${\cal O}(1/N^2)$ and does not contribute to the above calculations.
This graph is the main obstacle against attaining the charged
fixed point.

\section{Existence of the charged fixed point}

\subsection{Scaling \index{Scaling} near the charged fixed point}

By assuming the existence of the charged fixed point, it is
possible to derive exact scaling relations for the GL theory.\cite{deCalan2} 
For example, it is easy to prove that
if a charged fixed point exists then $\nu'=\nu$, where
$\nu'$ is the exponent of the penetration depth 
\index{Penetration depth} $\lambda$.
To this end, it is necessary to consider the GL model in
the ordered phase, that is, at $T<T_c$. In this situation
the vector potential becomes massive through the Higgs
mechanism,\cite{ZJ} with a mass $m_A=\lambda^{-1}$. 
The Higgs mechanism \index{Higgs mechanism} 
implies that there are no massless modes 
in the superconductor in the ordered phase. This shows  
a fundamental difference between a superconductor and a 
superfluid: The ordered phase of a superfluid contains 
a massless mode or Goldstone boson. 
The
Ginzburg parameter \index{Ginzburg parameter}
$\kappa$ is defined by the ratio 
between the Higgs mass and the vector potential mass:
$\kappa=m/m_A$. It can be shown that $m^2=u \rho_s/2$ and
$m_A^2=e^2\rho_s$, with $\rho_s$ being the superfluid
density. These formulas are easily obtained in a tree
level analysis of the Higgs mechanism. Their application
in the renormalized case follows from imposing
suitable renormalization conditions 
and the Ward identities.\cite{KleinNog1,Lee} 
Thus, $\kappa^2=u/2e^2=g/2f$.
Since
both $f$ and $g$ tend to a nonzero fixed point value in
the infrared ($m\to 0$), it follows that the
Ginzburg parameter $\kappa^2=\lambda^2/\xi^2=g/2f\to{\rm const}$
as $m\to 0$. As a consequence, the scaling
relation $\lambda\sim\xi$ holds and therefore the equality between the
corresponding critical exponents.

In the following it will be useful to use $m$ as the running
RG scale. Thus, $f=m^{d-4}Z_A e_0^2$ and $g=m^{d-4}Z_\psi^2 Z_g^{-1}u_0$.
Therefore,

\begin{equation}
\label{betaf3}
\beta_f\equiv m\frac{\partial f}{\partial m}=(\gamma_A+d-4)f,
\end{equation}
where

\begin{equation}
\label{gammaA}
\gamma_A\equiv m\frac{\partial\ln Z_A}{\partial m}.
\end{equation}
Thus, under the assumption of existence of the charged fixed
point we obtain that

\begin{equation}
\label{etaA}
\eta_A\equiv\gamma_A(f_*,g_*)=4-d.
\end{equation}
The above equation gives the {\it exact} 
anomalous dimension \index{Anomalous dimension} 
of the vector potential.\cite{Berg,Herbut} This means that the
large distance behavior at the critical point:

\begin{equation}
\langle A_\mu({\bf r})A_\mu({\bf r}')\rangle\sim
\frac{1}{|{\bf r}-{\bf r}'|^{d-2+\eta_A}}
\sim\frac{1}{|{\bf r}-{\bf r}'|^2},
\end{equation}
which holds for all $d\in(2,4]$.

It is instructive and experimentally relevant to compare the scaling
near a charged fixed point with the one governed by the $XY$ fixed
point. To this end, let us first note that the flow equation for $\kappa^2$
is written exactly as\cite{deCalan2}

\begin{equation}
\label{k}
m\frac{\partial\kappa^2}{\partial m}=\kappa^2\left(\frac{\beta_g}{g
}+4-d-\gamma_{A}\right).
\end{equation}
Since $\kappa^2=m^2/m_A^2$, it follows the following exact evolution
equation for $m_A^2$:\cite{deCalan2}

\begin{equation}
\label{mA}
m\frac{\partial m_{A}^2}{\partial m}=m_{A}^2\left(d-2+\gamma_{A}-
\frac{\beta_g}{g}\right).
\end{equation}
From Eq. (\ref{mA}) we obtain that near a fixed point $m_A$ behaves
as

\begin{equation}
\label{scalmA}
m_A\sim m^{(d-2+\eta_A)/2}.
\end{equation}
The above constitutes a rederivation of a result due to Herbut
and Tesanovic.\cite{Herbut}
For the charged fixed point $\eta_A=4-d$ and we obtain once more
that $\nu'=\nu$ (remember: $m=\xi^{-1}$ and $m_A=\lambda^{-1}$). For
the $XY$ fixed point, on the other hand, $\eta_A=0$ and we obtain
from Eq. (\ref{scalmA}) the scaling relation for the superconducting
$XY$ universality \index{XY universality class} 
class:\cite{Fisher}

\begin{equation}
\label{XYscal}
\nu'=\frac{\nu(d-2)}{2}.
\end{equation}

A further interesting consequence of the scaling relation
(\ref{scalmA}) is the following. The vector potential mass is
given in terms of the superfluid density $\rho_s$ as $m_A^2=e^2\rho_s$.
Eq. (\ref{betaf3}) implies the scaling $e^2\sim m^{\eta_A}$ near
the charged fixed point. Therefore, from Eq. (\ref{scalmA}) 
we obtain\cite{deCalan2}

\begin{equation}
\label{Josephson}
\rho_s\sim m^{d-2}\sim|t|^{\nu(d-2)},
\end{equation}
where $t$ is the reduced temperature. Eq. (\ref{Josephson}) is just
the Josephson relation\cite{Josephson} 
\index{Josephson relation} 
and we 
have thus shown that it is also valid
near the charged fixed point. In his original paper, Josephson has obtained
the above relation for the superfluid in the form
$\rho_s\sim t^{2\beta-\nu\eta}$.\cite{Josephson} Then he derived
the result (\ref{Josephson}) by assuming that hyperscaling holds, that
is, $d\nu=2-\alpha$, which together with the scaling relations
$\alpha+2\beta+\gamma=2$ and $\gamma=\nu(2-\eta)$ imply
$2\beta-\nu\eta=\nu(d-2)$. {\it We have obtained the result
(\ref{Josephson}) without using these supplementary scaling relations}.
Our result follows from the exact evolution equation for the vector
potential mass, Eq. (\ref{mA}). The form $\rho_s\sim t^{2\beta-\nu\eta}$
can be proven without any reference to the charge and is therefore
valid also for the superconductor. From this statement and Eq.
(\ref{Josephson}) we prove

\begin{equation}
2\beta-\nu\eta=\nu(d-2)
\end{equation}
for the superconductor.

Experiments in high quality crystals of $YBa_{2}Cu_{3}O_{7-\delta}$ (YBCO)
performed at zero field
verify very well Eq. (\ref{XYscal}) for the $d=3$ case.\cite{Kamal}
Most experiments are unable to probe the {\it charged} critical
region. However, in experiments involving critical dynamics the
situation is not clear. In this case we have again that different
scaling relations are obtained near the charged fixed point. In
general the AC conductivity \index{Conductivity} scales as
 $\sigma(\omega)\sim e^2\rho_s/(-i\omega)$. Since
$\rho_s\sim\xi^{2-d}$,
$e^2\sim\xi^{-\eta_A}$, and $\omega\sim\xi^{-z}$, where $z$ is the
dynamical critical exponent, we derive the scaling relation\cite{Nogueira}

\begin{equation}
\label{conduc}
\sigma(\omega)\sim\xi^{2-d+z-\eta_{A}}\sim|t|^{\nu(d-2-z-\eta_A)}.
\end{equation}
For the $XY$ universality class where $\eta_A=0$ we have
$\sigma(\omega)\sim\xi^{2-d+z}$.\cite{Fisher} Near the charged
fixed point, however, we obtain
$\sigma(\omega)\sim|t|^{\nu(2-z)}.$\cite{Nogueira,Mou}

\subsection{Duality \index{duality}}

Duality is a powerful tool in physics.\cite{KleinertBook}
It allows the mapping of a
weak coupling problem on a strong-coupling one. In the context of
statistical physics, it maps the low temperature expansion into
a high temperature expansion. In some cases, duality allows to
obtain exact information on the physical system. The classical example
is the two-dimensional Ising model, where the exact critical temperature
was obtained\cite{Kramers} before the exact solution appeared\cite{Onsager} 
The exact determination of the critical temperature
was possible because the Ising model has the self-duality property,
that is, the duality transformation has a fixed point. The self-duality
property is also verified in other systems and for $d>3$, like in
the $Z_2$ lattice gauge theories. However, the discreteness of the
gauge group makes these theories very similar from the point of
view of duality to the two dimensional Ising model. Self-duality
is more difficult to be found in continuous gauge groups. The
GL model, for example, has no such property, but as we shall see,
it is nevertheless almost self-dual.

In this section we shall discuss the field theoretical approach to
duality in the GL model. We shall show in detail how scaling works
in a disorder field theory \index{Disorder field theory} 
(DFT) for the superconducting phase transition.
The DFT to be discussed here was proposed first
by Kleinert nearly twenty years ago.\cite{Kleinert} This
formulation allowed to demonstrate that a tricritical point 
\index{Tricritical point}
exists
in the phase diagram of the superconductor. The existence of this
tricritical point allowed to build a consistent picture where the
strong-coupling limit -- which exhibits 
a second-order phase transition\cite{Dasgupta} -- and 
the weak coupling limit, with its weak first
order scenario, coexist with the normal phase, meeting at the
tricritical point. On the basis of the DFT, the
estimated value of $\kappa$ 
at the tricritical point was 
$\kappa_t\approx 0.8/\sqrt{2}$\cite{Kleinert} 
Early Monte
Carlo simulations\cite{Barth} gives, on the other hand, the
estimate $\kappa_t\approx 0.42/\sqrt{2}$. Remarkably, Kleinert's
estimate agree within $5$ \% with a recent, more precise, Monte
Carlo simulation by Mo, Hove, and Sudb{\o}.\cite{SUD}

The first scaling analysis of the DFT was made
by Kiometzis, Kleinert, and Schakel.\cite{Kiometzis} From the
analysis in Ref.~\refcite{Kiometzis} it was possible to establish
the value of the critical exponent $\nu$ as having a $XY$ value,
$\nu\simeq 0.67$. However, as we shall see, the scaling analysis
of the DFT contains some ambiguities which are not yet completely
solved.

\subsubsection{Duality in the lattice Ginzburg-Landau model
\index{Lattice Ginzburg-Landau model}}

A lattice version of the GL model has the Hamiltonian

\begin{equation}
\label{latticeGL}
H=-\beta\sum_{i,\mu}\cos(\nablab_\mu\theta_i-eA_{i\mu})
+\frac{1}{2}\sum_i(\mbox{\boldmath $\nabla$}\times{\bf A}_i)^2,
\end{equation}
where $\nabla_\mu$ is the lattice derivative,
$\nabla_\mu f_i\equiv f_{i+\hat{\mu}}-f_i$,
and $\beta=1/T$. The partition function is
then given by

\begin{equation}
Z=\int_{-\pi}^\pi\left[\prod_i\frac{d\theta_i}{2\pi}\right]
\int_{-\infty}^\infty\left[\prod_{i,\mu}dA_{i\mu}\right]
\exp(-H).
\end{equation}
The duality transformation \index{Duality transformation} 
can be done exactly when the Villain
form of the Hamiltonian is used. The Villain approximation\cite{Villain}
corresponds to the replacement

\begin{equation}
e^{\beta\cos x}\to \sum_{n=-\infty}^\infty
 e^{-\frac{\beta}{2}(x-2\pi n)^2},
\end{equation}
which turns out to be very accurate near the critical region.\cite{JK}

Assuming from now on the Villain approximation, 
\index{Villain approximation}   
we introduce an
auxiliary integer field $m_{i\mu}$ such that

\begin{eqnarray}
\label{transf}
\sum_{\{n_{i\mu}\}}\exp\left[-\frac{\beta}{2}(\nabla_\mu\theta_i
-eA_{i\mu}-2\pi n_{i\mu})^2\right]\nonumber\\
\propto \sum_{\{m_{i\mu}\}}
\exp\left[-\frac{1}{2\beta}m_{i\mu}^2+i(\nabla_\mu\theta_i
-eA_{i\mu})m_{i\mu}\right].
\end{eqnarray}
The proportionality factor above is not important in the following. 
All such proportionality factors will be neglected in the
foregoing manipulations. They correspond to smooth factors in the
temperature. Eq. (\ref{transf}) was obtained using the identity

\begin{equation}
\label{identity}
\sum_{m=-\infty}^\infty e^{(-t/2)m^2+ixm}=
\sqrt{\frac{2\pi}{t}}\sum_{n=-\infty}^{\infty}e^{(-1/2t)(x-2\pi n)^2}.
\end{equation}
The summation notation in Eq. (\ref{transf}) 
with a $\{n_{i\mu}\}$ means a {\it multiple} 
summation, analogous to multiple integration. 

By integrating out the angular variables $\theta_i$ we obtain the
partition function:

\begin{eqnarray}
\label{Z1}
Z&=&\int_{-\infty}^\infty\left[\prod_{i,\mu}dA_{i\mu}\right]
\sum_{\{{\bf m}_i\}}\delta_{\mbox{\scriptsize\boldmath $\nabla$}\cdot{\bf m}_i,0}
\exp\left\{\sum_i\left[-\frac{1}{2\beta}{\bf m}_i^2+ie {\bf A}_i\cdot
{\bf m}_i\right.\right.\nonumber\\
&-&\left.\left.\frac{1}{2}(\mbox{\boldmath $\nabla$}\times{\bf A}_i)^2
\right]\right\}.
\end{eqnarray}
The Kronecker delta constraint
$\mbox{\boldmath $\nabla$}\cdot{\bf m}_i=0$ generated by the
$\theta_i$ integrations implies that the links variables $m_{i\mu}$
form closed loops. These closed loops are interpreted as
magnetic vortices.\cite{KleinertBook} When $e=0$ the vector potential
decouples and we have, up to proportionality factor, the partition
function for the $XY$ model in terms of link variables:

\begin{equation}
\label{XYvort}
Z_{XY}=\sum_{\{{\bf m}_i\}}\delta_{\mbox{\scriptsize\boldmath $\nabla$}\cdot{\bf m}_i,0}
\exp\left(-\frac{1}{2\beta}\sum_i{\bf m}_i^2\right).
\end{equation}

We can solve the constraint on
${\bf m}_i$ by introducing a new integer link
variable through ${\bf m}_i=\mbox{\boldmath $\nabla$}\times{\bf l}_i$.
After integrating out ${\bf A}_i$ and using the Poisson
formula \index{Poisson formula}

\begin{equation}
\sum_{n=-\infty}^\infty F(n)=\sum_{m=-\infty}^\infty
\int_{-\infty}^\infty dx F(x) e^{2\pi imx}
\end{equation}
to go from the integer variables ${\bf l}_i$
to continuum variables ${\bf h}_i$, we obtain

\begin{eqnarray}
\label{dual}
Z&=&\int_{-\infty}^\infty\left[\prod_{i,\mu}dh_{i\mu}\right]
\sum_{\{{\bf m}_i\}}
\delta_{\mbox{\scriptsize\boldmath $\nabla$}\cdot{\bf m}_i,0}
\exp\left\{\sum_i\left[-\frac{1}{2\beta}(\mbox{\boldmath $\nabla$}
\times{\bf h}_i)^2
\right.\right.\nonumber\\
&-&\left.\left.\frac{e^2}{2}{\bf h}^2_i+2\pi i{\bf m}_i\cdot
{\bf h}_i\right]\right\}.
\end{eqnarray}
Eq. (\ref{dual}) corresponds to the dually transformed lattice
GL model.

By performing the ${\bf h}_i$ integration in Eq. (\ref{dual})
we obtain

\begin{equation}
\label{dual3}
Z=\sum_{\{{\bf m}_i\}}
\delta_{\mbox{\scriptsize\boldmath $\nabla$}\cdot{\bf m}_i,0}
\exp\left[-2\pi^2\beta\sum_{i,j,\mu}m_{i\mu}
G({\bf r}_i-{\bf r}_j)m_{j\mu}\right],
\end{equation}
where the Green function $G$ has the following behavior at
large distances:

\begin{equation}
G({\bf r}_i-{\bf r}_j)\sim
\frac{e^{-\sqrt{\beta} e|{\bf r}_i-{\bf r}_j|}}{4\pi|{\bf r}_i-{\bf r}_j|}.
\end{equation}
Thus, in the superconductor the magnetic vortex loops 
\index{Vortex loops}
interact with
a screened long range interaction.

Let us consider now the ``frozen'' superconductor limit\cite{Peskin}
of the dual representation (\ref{dual}). The ``frozen'' superconductor
corresponds to the zero temperature limit, $T\to 0$. In this case,
after integrating out ${\bf h}_i$, we obtain

\begin{equation}
\label{frozenSC}
Z_{\rm frozen}=\sum_{\{{\bf m}_i\}}
\delta_{\mbox{\scriptsize\boldmath $\nabla$}\cdot{\bf m}_i,0}\exp
\left(-\frac{2\pi^2}{e^2}\sum_i{\bf m}_i^2\right).
\end{equation}
Eq. (\ref{frozenSC}) has the same form as Eq. (\ref{XYvort}).
Thus, the ``frozen'' superconductor is the same as a $XY$ model
provided we identify

\begin{equation}
\label{Dirac}
e^2=\frac{4\pi^2}{T}.
\end{equation}
The above result allows us to localize a critical point on the
$e^2$-axis in the phase diagram in the $e^2-T$-plane. Using
Eq. (\ref{Dirac}) and the fact that $T_c\approx 3$ for the
$XY$ model in the Villain approximation,\cite{KleinertBook}
we obtain $e_c^2=4\pi^2/T_c\approx 13.159$ on the
$e^2$-axis. Thus, we can locate two limiting critical points
in the phase diagram, since we have in the $T$-axis the
Villain-$XY$ critical point at $T_c\approx 3$.

The vortex-vortex interaction in Eq. (\ref{dual3}) is singular at
short distances. Therefore, it is natural to introduce a vortex
core term with energy $\epsilon_0$ in the dual lattice Hamiltonian:

\begin{equation}
\label{dualH}
H_{\rm dual}=\sum_i\left[\frac{1}{2\beta}(\mbox{\boldmath $\nabla$}
\times{\bf h}_i)^2+\frac{e^2}{2}{\bf h}^2_i-2\pi i{\bf m}_i\cdot
{\bf h}_i+\frac{\epsilon_0}{2}{\bf m}_i^2\right].
\end{equation}
Using Eq. (\ref{identity}) and the integral representation

of the Kronecker delta

\begin{equation}
\label{intrepdelta}
\delta_{\mbox{\scriptsize\boldmath $\nabla$}\cdot{\bf m}_i,0}=
\int_{-\pi}^\pi\frac{d\theta_i}{2\pi}e^{i\theta_i(
\mbox{\boldmath $\nabla$}\cdot{\bf m}_i)},
\end{equation}
we obtain

\begin{equation}
\label{dualH1}
H_{\rm dual}'=\sum_i\left[\frac{1}{2\beta}(\mbox{\boldmath $\nabla$}
\times{\bf h}_i)^2+\frac{e^2}{2}{\bf h}^2\right]
+\sum_{i,\mu}
\frac{1}{2\epsilon_c}(\nabla_\mu\theta_i-2\pi i n_{i\mu}-2\pi h_{i\mu}
)^2.
\end{equation}
The Hamiltonian in Eq. (\ref{dualH1}) has the same form as the
lattice GL Hamiltonian in the Villain approximation, except that
in Eq. (\ref{dualH1}) the vector field is massive. Thus, we see
that there is almost a self-duality between them. When $e=0$
the Hamiltonian (\ref{dualH1}) is the dual of the (Villain)
lattice $XY$ Hamiltonian. Note that in this duality transformation
a locally gauge invariant model is mapped on a globally invariant one.

Let us set $e=0$ in Eq. (\ref{dualH}) and look for the phase diagram
in the $\beta-\epsilon_0$ plane. In such a phase diagram the point
$(\beta_c,0)$ corresponds to the $XY$ critical point. Integrating out
${\bf h}_i$ we obtain the partition function:

\begin{equation}
\label{XYdual}
Z'|_{e=0}=\sum_{\{{\bf m}_i\}}
\delta_{\mbox{\scriptsize\boldmath $\nabla$}\cdot{\bf m}_i,0}
\exp\left[-2\pi^2\beta\sum_{i,j,\mu}m_{i\mu}
\bar{G}({\bf r}_i-{\bf r}_j)m_{j\mu}-\frac{\epsilon_0}{2}
{\bf m}_i^2\right],
\end{equation}
where $\bar{G}$ behaves at large distances as

\begin{equation}
\bar{G}({\bf r}_i-{\bf r}_j)\sim
\frac{1}{4\pi|{\bf r}_i-{\bf r}_j|}.
\end{equation}
From Eqs. (\ref{XYvort}) and (\ref{XYdual}) we see that the point
$(0,1/2\beta_c)$ in the $\beta-\epsilon_0$ plane corresponds to an
``inverted'' $XY$ ($IXY$) 
\index{Inverted XY transition} 
transition.\cite{Dasgupta} By performing the
${\bf A}_i$ integration in Eq. (\ref{Z1}) we obtain

\begin{equation}
\label{Z2}
Z=\sum_{\{{\bf m}_i\}}
\delta_{\mbox{\scriptsize\boldmath $\nabla$}\cdot{\bf m}_i,0}
\exp\left[-\frac{e^2}{2}\sum_{i,j,\mu}m_{i\mu}
\bar{G}({\bf r}_i-{\bf r}_j)m_{j\mu}-\frac{1}{2\beta}
{\bf m}_i^2\right],
\end{equation}
and we see that the $IXY$ critical point corresponds
to $e_c^2=4\pi^2\beta_c$.\cite{Dasgupta}

\begin{figure}
\begin{picture}(60,152.05)
\unitlength.7mm
\put(20,3){\includegraphics[width=6cm]{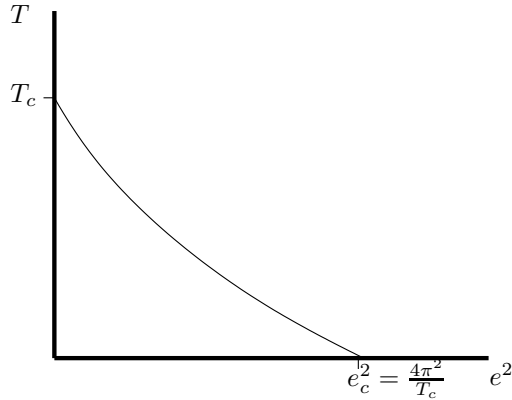}}
\put(105,0.3){$e^2$}
\put(78,0.1){$e_c^2=\frac{4\pi^2}{T_c}$}
\put(14,69){$T$}
\put(14,54){$T_c$}
\end{picture}
\caption{\label{PhaseDiag}
Schematic phase diagram showing the critical points
on the $e^2$- and $T$-axis. Here $T_c\approx 3$. The
ordered superconducting phase corresponds to
$0<e^2<e^2_c$.}
\end{figure}

Clearly the $IXY$ transition is a second-order phase transition.
Note that this transition arises in a lattice GL model where
the amplitude fluctuations are frozen (London limit). Therefore,
we should not expect to find a first-order phase transition in
this case. This London limit is appropriate when magnetic fluctuations
are relevant in the type II regime. In Fig. \ref{PhaseDiag} we
the approximate phase diagram.\cite{Kleinert}

\subsubsection{The disorder field theory 
\index{Disorder field theory}}

The $IXY$ universality class must have the same thermodynamic
exponents as the $XY$ model. For instance, we expect $\nu\simeq 0.67$
and, from the scaling analysis of Section 3.1, $\nu'=\nu$. This last
scaling relation has been confirmed in Monte Carlo simulations of
the lattice model (\ref{latticeGL}).\cite{Olsson} Although the
thermodynamic exponents are the same as in the $XY$ universality
class, critical exponents as $\eta$ and $\eta_A$ are not the
same. We have seen in Section 3.1 that $\eta_A=4-d$ near the charged
fixed point. The $d=3$ value, $\eta_A=1$, corresponds to the
$IXY$ universality class discussed in the preceding subsection. The
$XY$ universality class, on the other hand, has $\eta_A=0$. Also,
we have $-1<\eta<0$ in the $IXY$ universality class, while
$\eta>0$ in the $XY$ one.

Useful information from different universality classes and crossovers
in superconductors can be obtained from the DFT. The DFT is constructed
out the dual lattice Hamiltonians discussed in this section.
The {\it bare} DFT associated to the lattice Hamiltonian (\ref{dualH1})
is given by\cite{GFD,Kiometzis}
\begin{equation}
\label{DFT}
{\cal H}_{\rm DFT}=\frac{1}{2}(\mbox{\boldmath $\nabla$}\times{\bf h}_0)^2
+\frac{m_{A,0}^2}{2}{\bf h}_0^2+|(\mbox{\boldmath $\nabla$}-i\tilde{e}_0
{\bf h}_0)\phi_0|^2+\tilde{m}_0^2|\phi_0|^2+\frac{\tilde{u}_0}{2}|\phi_0|^4,
\end{equation}
where the bare 
{\it dual charge} \index{Dual charge}
$\tilde{e}_0\equiv 2\pi m_{A,0}/e_0$. $m_{A,0}$ 
is the bare mass of the vector potential in the original theory. We see
that the disordered phase of the DFT corresponds to the ordered phase of
the GL model. This is a general feature of all duality transformations:
the low temperature phase is mapped in the high temperature phase.
This justifies the denomination ``disorder field theory'' for the
continuum limit of the lattice dual model. The field $\phi_0$ is
the bare {\it disorder parameter field}. In the superconducting
phase $\langle\phi_0\rangle=0$, while the {\it order parameter}
in the original GL model $\langle\psi_0\rangle\neq 0$. Conversely,
the normal phase corresponds to $\langle\phi_0\rangle\neq 0$ and
$\langle\psi_0\rangle=0$.

It should be noted that the Hamiltonian (\ref{DFT}) is a generalization
of the London model. The field ${\bf h}_0$ is in fact the magnetic induction,
while $|\phi_0|^2$ gives the vortex loop density.

The renormalization of the Hamiltonian (\ref{DFT}) is similar to the one of
the GL model, up to the following subtlety. From the Ward identities we
obtain that the mass term for the vector field is not renormalized, that
is, $m_{A,0}^2{\bf h}_0^2/2=m_A^2{\bf h}^2/2$, where the absence of
the zeroes subindices indicate renormalized quantities. Since the
renormalized induction field is given by ${\bf h}=Z_h^{-1/2}{\bf h}_0$, we
obtain

\begin{equation}
\label{mArenorm}
m_A^2=Z_h m_{A,0}^2.
\end{equation}
The dual charge \index{Dual charge} renormalizes in a similar way:

\begin{equation}
\label{dualcrenorm}
\tilde{e}^2=Z_h\tilde{e}_0^2.
\end{equation}
Since $\tilde{e}_0^2=4\pi^2 m_{A,0}^2/e_0^2$, it follows from Eqs.
(\ref{mArenorm}) and (\ref{dualcrenorm}) that the Cooper pair charge $e_0$
is not renormalized in the DFT, $e=e_0$.

It is important to remark that the vector potential mass renormalization
in the DFT involves only one renormalization constant, while the
same is not true in the GL model.\cite{KleinNog1}

Due to the presence of a massive vector field, the DFT has a ambiguous
scaling.\cite{deCalan2}
This ambiguity was a matter of debate recently
\cite{Herbut1,HComment,KSComment} Let's see how it works. 
In Ref.~\refcite{Kiometzis} 
the scaling chosen was in principle very natural,
with the bare masses behaving in the same way with mean
field exponents: $\tilde{m}_0^2\sim|t|$ and $m_{A,0}^2\sim|t|$.
Renormalization is employed as usual. We have,

\begin{equation}
\label{dmass}
\tilde{m}_0^2=Z_{\tilde{m}} Z_\phi^{-1}\tilde{m}^2,
\end{equation}
and therefore,

\begin{equation}
\label{flowdm}
\tilde{m}\frac{\partial\tilde{m}_0^2}{\partial\tilde{m}}=
(2+\gamma_{\tilde{m}})\tilde{m}_0^2,
\end{equation}
where $\gamma_{\tilde{m}}$ is defined in a way similar to 
$\gamma_m$ in Eq. (\ref{gammam}). Since
$\tilde{m}_0^2/m_{A,0}^2={\rm const}$, we obtain

\begin{equation}
\label{flowmA0}
\tilde{m}\frac{\partial m_{A,0}^2}{\partial\tilde{m}}=
(2+\gamma_{\tilde{m}})m_{A,0}^2.
\end{equation}
Let us define the dimensionless renormalized dual coupling
by $\tilde{f}=\tilde{e}^2/\tilde{m}$.
From Eqs. (\ref{dualcrenorm}) and (\ref{flowmA0}), we obtain
the $\beta$-function:

\begin{equation}
\label{betadf}
\beta_{\tilde{f}}\equiv\tilde{m}\frac{\partial\tilde{f}}{
\partial\tilde{m}}=(\gamma_h+\gamma_{\tilde{m}}+1)\tilde{f},
\end{equation}
where

\begin{equation}
\gamma_h\equiv\tilde{m}\frac{\partial\ln Z_h}{\partial\tilde{m}}.
\end{equation}
Now, we can easily obtain the following bound

\begin{equation}
\label{bound}
\gamma_{\tilde{m}}+1\leq\frac{1}{\nu}-1\leq 1.
\end{equation}
It is also straightforward to show that $\gamma_h\geq 0$. Therefore,
the infrared stable fixed point to (\ref{betadf}) is at $\tilde{f}_*=0$.
This implies that the critical exponent $\nu$ has a $XY$ value and
that $\eta_h\equiv\gamma_h^*=0$.

From Eqs. (\ref{mArenorm}) and (\ref{flowmA0}) we obtain

\begin{equation}
\label{evolmAdual}
\tilde{m}\frac{\partial m_A^2}{\partial\tilde{m}}=
(\gamma_h+\gamma_{\tilde{m}}+2)m_A^2.
\end{equation}
Near the fixed point the above equation becomes

\begin{equation}
\tilde{m}\frac{\partial m_A^2}{\partial\tilde{m}}
\approx\frac{1}{\nu}m_A^2,
\end{equation}
which implies the scaling

\begin{equation}
m_A^2\sim\tilde{m}^{1/\nu}.
\end{equation}
Since $\tilde{m}\sim|t|^\nu$ the above scaling implies that
the penetration depth exponent is given exactly by
$\nu'=1/2$.

Another possible scaling is the one considered by 
Herbut\cite{Herbut1} where it is assumed that $m_{A,0}^2={\rm const}$.
Within this scaling we obtain instead Eq. (\ref{betadf}) the
following $\beta$-function:

\begin{equation}
\label{betadf1}
\beta_{\tilde{f}}=(\gamma_h-1)\tilde{f},
\end{equation}
which is similar to the $\beta$-function of the coupling $f$
in the GL model at $d=3$. From Eq. (\ref{mArenorm}) we
obtain

\begin{equation}
\label{evolmAdual1}
\tilde{m}\frac{\partial m_A^2}{\partial\tilde{m}}=\gamma_h m_A^2.
\end{equation}
The $\beta$-function for the coupling
$\tilde{g}=\tilde{u}/\tilde{m}$ contains functions of the ratio
$\tilde{m}/m_A$ multiplying every power of $\tilde{f}$.\cite{Herbut1}
Due to the evolution equation (\ref{evolmAdual1}), we see that
$m_A^2\sim\tilde{m}^{\gamma_h^*}\sim\tilde{m}$. Therefore,
$\tilde{m}/m_A\to 0$ as the critical point is approached. Thus,
fixed point $\tilde{g}_*$ is the same as in the $XY$ model and
once more the critical exponent $\nu$ has a $XY$ value.
\cite{Herbut1} However, from the exact scaling behavior
$m_A^2\sim\tilde{m}$ we see that the penetration depth
exponent is given by $\nu'=\nu/2$, which corresponds
to the same value as in the 3D $XY$ superconducting
universality class. Therefore, the scaling considered in
Ref.~\refcite{Herbut1} does not give the expected value
for the $IXY$ universality class, which should be
$\nu'=\nu\approx 2/3$.\cite{Herbut} It was shown in
Ref.~\refcite{deCalan2} that the $IXY$ universality class can only
be obtained if the bare mass $m_{A,0}^2\sim|t|^{\zeta}$, where
$\zeta=2\nu\approx 4/3$, i.e., $m_{A,0}^2$ should scale as
$m^2$ of the original GL model. It seems that if we want that
the dual model implies the result $\nu'=\nu$, we have to
make this assumption.

An alternative scenario for the scaling behavior in the
DFT is the following. We have shown that the $XY$ model
dualizes in a GL model. The GL model, on the other hand,
dualizes on the DFT whose Hamiltonian is given in Eq. (\ref{DFT}).
Now, the dual of the dual must be of course the original model.
This means that the GL model should dualize in a $XY$ model and
therefore the DFT Hamiltonian should be equivalent to it. On the
basis of this argument we should expect a scaling consistent
with the $XY$ universality class, instead of $IXY$. If we
accept this argument, we are led to the conclusion that the
correct scaling behavior should assume $m_{A,0}^2={\rm const}$
as in Ref.~\refcite{Herbut1} to obtain the $XY$ scaling of the
penetration depth, $\nu'=\nu/2$.

\section{The physical meaning of the critical exponent $\eta$}

In the superconducting phase transition only the exponents
$\nu$, $\nu'$, and $\alpha$ are measured.
Here $\alpha$ is the specific heat exponent, which is related 
to $\nu$ by the hyperscaling relation $d\nu=2-\alpha$.  
At present the critical
exponent $\eta$ is not measured and we can even doubt of its
physical significance. We can argue that the superconducting
order parameter cannot be considered to be a physical measurable
quantity because its conjugate field has no physical meaning.
In a ferromagnet the field conjugate to the magnetization is just
the external magnetic field, which can be controlled by
experiments. Another problem is that the order parameter
$\langle\psi\rangle$ is not gauge invariant. Thus, a
calculation of $\langle\psi({\bf r})\psi^\dagger({\bf r}')\rangle$
will depend on the gauge choice and, as a consequence, $\eta$ will
also be gauge dependent. \index{Gauge dependent}

In this Section we shall show that it is possible to
define a gauge-independent $\eta$ exponent and discuss its
physical significance. The physical meaning of $\eta$
arises due to a special feature which at first glance looks very much
like a pathology: {\it it has a negative sign}. Indeed, we would
expect from very general non-perturbative arguments that $\eta$
should be positive. In the case of a pure $|\psi|^4$ theory we
can prove that $\eta\geq 0$ in the following way. In momentum
space the correlation function
$G({\bf r}-{\bf r}')\equiv\langle\psi({\bf r})\psi^\dagger({\bf r}')\rangle$
has the spectral representation:\cite{GJ}

\begin{equation}
\label{KL}
\hat{G}({\bf p})=\int_0^\infty d\mu \frac{\rho(\mu)}{{\bf p}^2+\mu^2},
\end{equation}
where the spectral weight $\rho(\mu)$ satisfies the sum rule

\begin{equation}
\label{sr}
\int_0^\infty d\mu\rho(\mu)=1.
\end{equation}
The above representation is well known in quantum field theory and
is called K\"allen-Lehmann spectral 
\index{K\"allen-Lehmann representation}
representation.\cite{GJ}
Now, because of the condition (\ref{sr}) on the spectral weight,
we have the inequality

\begin{equation}
\label{ineq}
\hat{G}({\bf p})\leq\frac{1}{{\bf p}^2}.
\end{equation}
If one assumes the low momentum behavior at the critical point
$\hat{G}({\bf p})\sim 1/|{\bf p}|^{2-\eta}$, we obtain from
the inequality (\ref{ineq}) that $\eta\geq 0$.

In the case of the GL model, all calculations of $\eta$ give a
negative value in the interval $-1<\eta<0$ for $d=3$.
\cite{HLM,Herbut,Folk,Berg,Radz,Hove,Olsson1,KleinNog1} In
general it is argued that since
$\hat{G}({\bf p})\sim 1/|{\bf p}|^{2-\eta}$, we have in real
space the large distance behavior at the critical point:

\begin{equation}
\label{G}
G({\bf r}-{\bf r}')\sim\frac{1}{|{\bf r}-{\bf r}'|^{d-2+\eta}}.
\end{equation}
The above will not diverge as $|{\bf r}-{\bf r}'|\to\infty$
provided $\eta>2-d$ and for this reason we could have in
principle a negative $\eta$ exponent. Such an argument is
certainly not correct in the case of pure $|\psi|^4$ theory
where the K\"allen-Lehmann representation holds which
implies $\eta\geq 0$. In the case of the GL model the situation is
much more subtle and the K\"allen-Lehmann representation does
not apply, at least not in the above form.\cite{Nogueira,Mo}

In order to give a physical interpretation to the negative sign
of $\eta$ in superconductors, let us consider a one-loop
approximation at the critical point and fixed dimensionality $d=3$.
The calculation is uncontrolled but serves to illustrate the
main point. Assuming $N=2$,
the vector potential propagator is then given by

\begin{equation}
\label{D}
D_{\mu\nu}({\bf p})=\frac{1}{{\bf p}^2+{e^2}|{\bf p}|/16}
\left(\delta_{\mu\nu}-\frac{p_\mu p_\nu}{{\bf p}^2}\right),
\end{equation}
while the order parameter two-point correlation function is
 \begin{equation}
\label{G1}
G({\bf p})=\frac{1}{{\bf p}^2-{e^2}|{\bf p}|/4}.
\end{equation}
Now, we see from Eq. (\ref{D}) that as $|{\bf p}|\to 0$ the second
term in the denominator dominates implying $\eta_A=1$. However,
the same argument does not apply to Eq. (\ref{G1}) because the
second term in the denominator has a negative sign in front of it and
therefore the ${\bf p}^2$ term is still relevant. Thus, there is a
momentum space instability in the problem similar to the one
encountered in theories of magnetic systems exhibiting a
Lifshitz \index{Lifshitz point}
point.\cite{Diehl} There the Hamiltonian
already contains the momentum space instability from the very
beginning due to the presence of higher order derivatives.
\cite{Diehl} Due to this, the susceptibility in those
magnetic systems has a maximum at
a nonzero value of ${\bf p}$. This lead to the appearance of a
modulated regime in the phase diagram, which is plotted in
the $P-T$-plane, where $P$ is the ratio between two competing
interactions. It was conjectured
in Ref.~\refcite{Nogueira} that a similar behavior occurs in
the superconductor. The modulated regime would be associated to
the type II behavior, which would be in this way analogous to
the helical phase in magnetic systems exhibiting a Lifshitz
point. In the case of magnetic systems, the Lifshitz point
is the point in the phase diagram where the paramagnetic,
ferromagnetic, and helical phases coexist. In the case of the
superconductor it corresponds to the point where the type I, type II,
and normal phases coexist. In straight analogy with magnetic
systems, the phase diagram is plotted in the $\kappa^2-T$-plane.
Note that in the case of the GL model the Lifshitz point-like
behavior would be generated by thermal fluctuations.

Further insight in this problem
can be obtained by looking at the propagator in
the $1/N$ expansion already discussed in this review. The
self-energy at the critical point and ${\cal O}(1/N)$ is given by

\begin{equation}
\label{se1/N}
\Sigma({\bf p})=\frac{40}{\pi^2 N}{\bf p}^2\ln
\left(\frac{|{\bf p}|}{Ne^2}\right).
\end{equation}
Thus, besides the pole at ${\bf p}=0$ we have also a pole
at

\begin{equation}
\label{pole}
|{\bf p}_0|=Ne^2\exp\left(-\frac{\pi^2N}{40}\right).
\end{equation}
This instability is similar to the one leading to chiral symmetry
breaking in three-dimensional QED (QED3).
\cite{QED3} The difference is that in QED3 the instability occurs
with respect to the mass, which is dynamically generated by
spontaneous chiral symmetry breaking. Thus, if $M$ is the generated
fermion mass, the pole of the propagator occurs at\cite{QED3}

\begin{equation}
\label{M}
M=Ne^2\exp\left(-\frac{\pi^2N}{8}\right).
\end{equation}

We have not yet discussed how to cure the disease which results from
the lack of gauge invariance \index{Gauge invariance}
of the order parameter
correlation function $G({\bf r}-{\bf r}')$. Our physical
interpretation of the critical exponent $\eta$ given above has no
value if $\eta$ is a gauge-dependent quantity. The best thing to do
is to define a gauge-invariant correlation function for the
order parameter. The choice of such a gauge-invariant
correlation function is not unique but as we shall show, there is
one whose value of $\eta$ coincides with the one which is obtained
by computing it
in the Coulomb gauge, which is the gauge we are using in this
review.

A popular gauge-invariant correlation function 
\index{Gauge-invariant correlation function} 
which is
often used in the literature is
\begin{equation}
\label{g-invG1}
{\cal G}({\bf r}-{\bf r}')=
\left\langle\psi({\bf r})\exp\left[-ie\int_{\bf r}^
{{\bf r}'} d{\bf r}''\cdot{\bf A}({\bf r}'')\right]
\psi^\dagger({\bf r}')\right\rangle.
\end{equation}
A calculation of $\eta$ using the above correlation function
was carried out recently by Kleinert and Schakel.\cite{KS}
They calculated this exponent using both the
$\epsilon$-expansion and the $1/N$-expansion.
In the former
case the result is

\begin{equation}
\label{g-inv-eta1}
\eta=-\frac{36}{N}\epsilon,
\end{equation}
while in the latter case $\eta$ was computed for
arbitrary dimensionality $d\in(2,4)$ and up to
order $1/N$\cite{KS}

\begin{equation}
\label{g-inv-eta2}
\eta=-\frac{1}{N}\frac{(d^2+2d-6)\Gamma(d-2)}{\Gamma(2-d/2)\Gamma^2
(d/2-1)\Gamma(d/2)}.
\end{equation}
By setting $d=4-\epsilon$ in Eq. (\ref{g-inv-eta2}) and expanding
to order $\epsilon$ we obtain Eq. (\ref{g-inv-eta1}). In a
GL Hamiltonian with a gauge-fixing term

\begin{equation}
\label{gfixing}
{\cal H}_{\rm gf}=\frac{1}{2\alpha}(\nabla\cdot{\bf A})^2,
\end{equation}
the above expressions of $\eta$ corresponds to values that
would have been obtained by fixing the gauge
$\alpha=-3$ in the $\epsilon$-expansion case and
$\alpha=1-d$ in the case of the $1/N$-expansion. Thus, in each
case the value of $\eta$ does not agree with the one that is
obtained by fixing the Coulomb gauge, which corresponds
to $\alpha=0$. Note, however, that the above gauge-independent
results both confirm that $\eta$ is indeed negative.

\begin{figure}
\begin{picture}(60,152.05)
\unitlength.7mm
\put(20,3){\includegraphics[width=6cm]{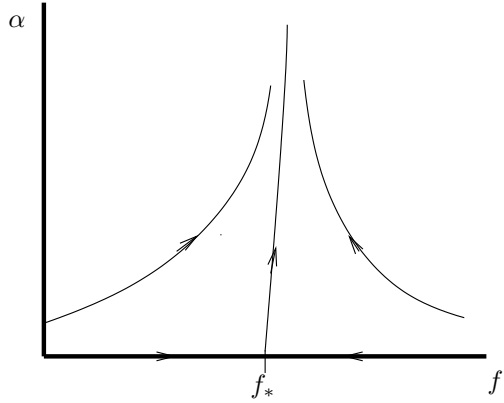}}
\put(105,0.3){$f$}
\put(60,-0.4){$f_*$}
\put(14,69){$\alpha$}
\end{picture}
\caption{\label{flow-gf}
Schematic flow diagram in the $f-\alpha$-plane.}
\end{figure}

A different point of view discussed
in Ref.~\refcite{deCalan} focus instead in the flow of
the gauge-fixing parameter $\alpha$. From the Ward
identities it follows that the gauge-fixing parameter 
\index{Gauge-fixing parameter} 
renormalizes as

\begin{equation}
\label{renorm-gf}
\alpha=Z_A^{-1}\alpha_0,
\end{equation}
which implies the $\beta$-function

\begin{equation}
\label{beta-gf}
\beta_\alpha=-\gamma_A \alpha.
\end{equation}
Since at the charge fixed point we have $\gamma_A(f_*,g_*)=4-d$,
the only way to get a fixed point to Eq. (\ref{beta-gf}) when
$d\in (2,4)$
is to set $\alpha=0$. Due to the negative sign
in Eq. (\ref{beta-gf}), the flow is unstable for
arbitrary nonzero $\alpha$. Thus, it is clear that stable
charged fixed points can be obtained only if the Coulomb
gauge corresponds to the fixed point of the theory
in $2<d<4$. For $d=4$, which is the case of interest for
particle physicists, any value of $\alpha$ can be chosen, since
in this case $\gamma_A(f_*,g_*)=0$.
In Fig. \ref{flow-gf} we show a schematic flow diagram in the
$f-\alpha$-plane for the case where $d\in(2,4)$.
Based on this scenario, we
are led to conclude that the above gauge-independent results
for $\eta$ do not correspond to the infrared stable fixed
point. This means that the correlation function (\ref{g-invG1})
is not a good choice of gauge-invariant correlation function giving
a gauge-independent value of the $\eta$ exponent. In fact,
the correlation function (\ref{g-invG1}) even fails to lead
to long-range order: It can be rigorously shown that in the
dimensions of interest the correlation function
(\ref{g-invG1}) always decay to zero.\cite{Frohlich} A better
gauge-invariant correlation function is\cite{KK}

\begin{equation}
\label{g-invG2}
{\cal G}({\bf r}-{\bf r}')=
\left\langle\psi({\bf r})\exp\left[-ie\int
d^d{\bf r}''{\bf A}({\bf r}'')\cdot{\bf b}({\bf r}'') \right]
\psi^\dagger({\bf r}')\right\rangle,
\end{equation}
where

\begin{equation}
{\bf b}({\bf r}'')=\nablab V({\bf r}''-{\bf r})
-\nablab V({\bf r}''-{\bf r}'),
\end{equation}
with

\begin{equation}
-\nabla^2V({\bf r})=\delta^d({\bf r}).
\end{equation}
Since the above correlation function is gauge-invariant, any
gauge can be fixed to calculate it. The end result will be
always gauge-independent.
It is easy to see that in the Coulomb gauge

\begin{equation}
{\cal G}({\bf r}-{\bf r}')|_{\rm Coulomb}=
\langle\psi({\bf r})\psi^\dagger({\bf r}')\rangle |_{\alpha=0}.
\end{equation}
Therefore in such a
scenario the $\eta$ exponent corresponds precisely
to the one we have calculated previously in the Coulomb gauge.
Furthermore, it can be shown that the correlation
function (\ref{g-invG2}) exhibits long-range order,\cite{KK}
in contrast to the one given in Eq. (\ref{g-invG1}).

\section{Renormalization group \index{Renormalization group} 
calculation at fixed dimension
and below $T_c$}

In the $\epsilon$-expansion RG approach the same RG functions
are obtained no matter the calculation is carried out below or
above $T_c$. As dictated by the Ward identities the singular
behavior is exactly the same above and below $T_c$.\cite{ZJ}
However, if the calculation is done in fixed dimension
$d=3$ and below $T_c$ the situation is different and the RG
functions depend explicitly on the Ginzburg parameter $\kappa$.
We shall not give the details of this approach here. Instead, we
shall concentrate on the physical aspects of this new approach which
allows to obtain a charge fixed point 
\index{Charged fixed point}
at one-loop order. The
interested reader is referred to Ref.~\refcite{KleinNog1} for the
technical details.

The approach we are going to discuss is not really perturbative.
Actually, only the powers of $f$ are being effectively counted and
the powers of $g$ are counted only partially. Thus, by one-loop
we mean first-order in $f$. The point is that $\kappa$ arises
in the calculations in two different ways: as the ratio
between the masses $\kappa=m/m_A$ and as the ratio between
coupling constants, $\kappa^2=g/2f$. The coupling $g$, when it
appears, is eliminated in favor of $\kappa$ and the RG flow
is in this way 
parametrized in terms of $\kappa$ and $f$. This way of working
is of course more complicated than more usual RG approaches but
it has the advantage of being physically more appealing due to the
explicit presence of $\kappa$ in the RG functions. In the classical
Abrikosov solution of the GL model in an external magnetic field
$\kappa$ appears explicitly and the existence of two types of
superconductivity is made evident.\cite{Fetter} For instance,
the slope of the magnetization curve near $H_{c2}$ is given
by

\begin{equation}
\frac{dM}{dH}=\frac{1}{4\pi\beta_A(2\kappa^2-1)},
\end{equation}
where $\beta_A$ is the Abrikosov parameter. The above expression is
singular at $\kappa=1/\sqrt{2}$, which corresponds to the point
separating type I from type II superconductivity. Such a
singular behavior at $\kappa=1/\sqrt{2}$ should be also visible
in the GL model with a thermally fluctuating vector potential.
The new approach introduced in Ref.~\refcite{KleinNog1} makes this
feature explicit in a RG context. As we shall see, this aspect of this new
approach is crucial to obtain the charged fixed point at
$d=3$ and $N=2$.

The only RG function that is singular at
$\kappa=1/\sqrt{2}$ is $\gamma_A$:

\begin{equation}
\label{gammaA1loop}
\gamma_A=\frac{\sqrt{2}C(\kappa)f}{24\pi(2\kappa^2-1)^3},
\end{equation}
where

\begin{equation}
C(\kappa)=4\kappa^6+10\kappa^4-24\sqrt{2}\kappa^3
+27\kappa^2+4\sqrt{2}\kappa-{1}/{2}.
\end{equation}
The $\beta$-function for $\kappa^2$ is given by

\begin{equation}
\label{betak2}
\beta_{\kappa^2}=(2\gamma_\pi-\gamma_A-\zeta_\pi)\kappa^2,
\end{equation}
where

\begin{equation}
\label{gammapi1loop}
\gamma_\pi=\frac{\kappa\,f}{12\pi}\frac{2 \kappa ^2+ \sqrt{2} \kappa -8
}{ \sqrt{2} \kappa +1},
\end{equation}
\begin{equation}
\label{zetapi}
\zeta_\pi
=-\frac{\sqrt{2}}{4\pi}f\left(\frac{3\kappa^2}{2}+
\frac{1}{\sqrt{2}\kappa}\right).
\end{equation}
As before, the charged fixed point at $d=3$ is determined by the
condition $\gamma_A(f_*,\kappa_*)=1$. This leads to the
fixed points

\begin{equation}
f_*\approx 0.3, ~~~~~~\kappa_*\approx 1.17/\sqrt{2}.
\label{@fpvs}\end{equation}
Note that $\kappa_*$ is slightly above the value $1/\sqrt{2}$ and
therefore the charged fixed point occurs in the type II regime.

The reason why a charged fixed point is obtained in the
above analysis is similar to the reason why the $1/N$-expansion
leads to a charged fixed point already at order $1/N$: The
fixed point coupling $f_*$ is small enough such that
a $f_*^2$-term is strongly suppressed in the other RG functions.
It is a large $f^2$-term in $\beta_g$ that spoils at $N=2$ the charged
fixed point in the HLM theory. In order to explain why this new
method is so successful, let us define an effective coupling
$\bar{f}$ by

\begin{equation}
\label{fbareq}
\gamma_A(\bar{f},\kappa)=1.
\end{equation}
The above equation defines a critical line in the sense that
the $\beta_f$ vanishes on this line. Note, however, that
$\beta_{\kappa^2}$ does not vanish in general. From
Eq. (\ref{fbareq}) we obtain

\begin{equation}
\label{fbar}
\bar{f}(\kappa)=\frac{24\pi(2\kappa^2-1)^3}{\sqrt{2}C(\kappa)}.
\end{equation}
From the above equation we see that $\bar{f}(\kappa)$ becomes
very small when $\kappa$ approaches $1/\sqrt{2}$ from the right.
Precisely at $\kappa=1/\sqrt{2}$ we have $\bar{f}=0$. This
behavior suggests that the best approximation scheme should be
one where the small parameter is given by
$\Delta\kappa\equiv\kappa-1/\sqrt{2}$. Now it is easy to see
that the $\epsilon$-expansion based RG fails because it
effectively expands
around $\kappa=0$ and therefore it corresponds to the deep type I
regime where the transition is clearly first-order. Furthermore,
$C(\kappa)$ vanishes at $\kappa=0.096/\sqrt{2}$ and
$\bar{f}$ becomes very large as this value of $\kappa$ is
approached from the left. Thus, a perturbation expansion
around $\kappa=0$ breaks down at $\kappa=0.096/\sqrt{2}$. There
is an ``infinite barrier'' separating the deep type I from the type II
regime. In the interval $0.096/\sqrt{2}<\kappa<1/\sqrt{2}$ the
effective coupling $\bar{f}$ is negative and thus unphysical.
The coupling $\bar{f}$ can be really small only for $\kappa>1/\sqrt{2}$,
i.e., in the type II regime.

Note that our one-loop approximation gives only one charged
fixed point. The tricritical fixed point is absent in this
approximation. This behavior also occurs in the $1/N$-expansion
where only one charged fixed point is found. A higher order
calculation is necessary to obtain the tricritical fixed point.
At two loops the singular behavior in $\kappa$ is expected to
change. Thus, instead finding a singularity at $\kappa=1/\sqrt{2}$,
which is the same as in the mean-field solution, we expect to
find a singular behavior at $\kappa_t\approx 0.8/\sqrt{2}$,
in agreement with Refs.~\refcite{Kleinert} and \refcite{SUD}.

\section{Concluding remarks}

In this paper we have reviewed several modern field-theoretic approaches
in the superconducting phase transition. We have emphasized some special
topics which are not extensively discussed in the literature.
In particular, the scaling behavior of the continuum dual model was
analysed in more detail than in the original publications.
The duality scenario is physically
and conceptually very important, but its scaling behavior
 is not yet fully understood. Another topic that deserves
further attention
is the recently conjectured Lifshitz point-like
behavior in the GL model.\cite{Nogueira,KleinNog1} Such a scenario
provides an interesting possibility
to understand  physically
the
negative value
of the
critical exponent $\eta$.

\section*{Acknowledgments}
\addcontentsline{toc}{section}{Acknowledgments}

This review is based on lectures delivered by F.S.N. at the
``Ising Lectures 2002'' at the Institute for Condensed Matter
Physics (ICMP), Lviv, Ukraine. F.S.N. is grateful
to Yurij Holovatch for the invitation to Lviv and the hospitality
of the ICMP. F.S.N. also thanks
 the
Alexander von Humboldt Foundation for the financial support.


\begin{thebibliography}{8.}
\bibitem{KleinertBook} A detailed discussion on duality can be
found in the textbook Ref.~\refcite{DO}.
\end{thebibliography}
\end{document}